\documentclass{jpsj-suppl}
\usepackage{txfonts} %Please comment out this line unless the txfonts package is availabe in your LaTeX system.

\title{Hadronic EDM and New physics beyond standard model}

\author{Nodoka \textsc{Yamanaka}$^{1}$}

\inst{$^{1}$iTHES Research Group, RIKEN, 2-1 Hirosawa, Wako, Saitama 351-0198, Japan}

\email{nodoka.yamanaka@riken.jp}

\recdate{June 15, 2016}

\abst{
The nuclear electric dipole moment is a very sensitive probe of CP violation beyond the standard model, and for light nuclei, it can be evaluated accurately using the few-body calculational methods. 
In this talk, we present the result of the evaluation of the electric dipole moment of 3-body and 4-body systems using the Gaussian expansion method in the ab initio approach and in the cluster model. 
We also give the future prospects for the discovery of new physics beyond it within the sensitivity of prepared experiments.
}

\kword{Electric dipole moment, CP violation, nuclear cluster structure}

\begin{document}
\maketitle

\def\vc#1{\mbox{\boldmath $#1$}}

\section{Introduction}

Our Universe is filled of matter, but it is believed that the matter and antimatter numbers had almost no asymmetry in the very early time.
The generation of this asymmetry is explained by the CP violation at the fundamental level \cite{sakharov}, but it is currently known that the standard model of particle physics cannot sufficiently make our Universe matter abundant.
The search for CP violation beyond the standard model is now one of the most important goal of particle physics.

The electric dipole moment (EDM) of nuclei is known to be a very sensitive probe of new physics beyond standard model \cite{khriplovichbook,pospelovreview,edmreview,edmreview2,yamanakabook}.
The nuclear EDM has several advantages, such as the accurate measurability in experiment \cite{storage1,storage2}, the small standard model contribution \cite{ckm,yamanakasmedm,yamanakasmedmcc}, and the absence of atomic electrons which screen the EDM of the constituents \cite{schiff}.
Recently, the experimental measurement of the EDM of light nuclei using storage rings is in preparation \cite{storage}, and also much theoretical development has been achieved.

Theoretically, one of the most attractive feature of the nuclear EDM is that it can accurately be evaluated using few-body methods.
Using the Gaussian expansion method \cite{hiyama}, a powerful few-body method, it is possible to evaluate the EDM of light nuclei, such as the deuteron, $^3$He, and $^3$H.
Moreover, we can also evaluate the EDM of $^6$Li, $^9$Be, and $^{13}$C in the cluster model, which is known to well describe the low energy nuclear structure \cite{clusterreview3}.
In this talk, we present the theoretical evaluations of the nuclear EDM and their results.

In the next section, we give the nuclear interactions and present the Gaussian Expansion Method.
We then define the EDM, which is generated by two leading CP violating processes.
In Section \ref{sec:results}, we show the result of the theoretical calculations of the EDM of light nuclei and analyze it.
In Section \ref{sec:prospects}, we give the prospects for the discovery of new physics beyond standard model through nuclear EDM experiments.
We finally summarize this discussion in Section \ref{sec:summary}.

\section{Interactions and methodology}

\subsection{The Bare $N-N$ interaction}

\begin{figure}[tbh]
\begin{center}
\includegraphics[width=8cm]{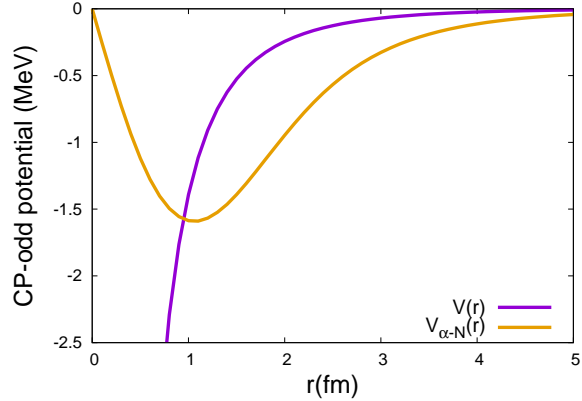}
\caption{
The radial dependence of the CP-odd nuclear force $V (r)$ and the CP-odd $\alpha - N$ interaction $V_{\alpha - N} (r)$, obtained after folding.
}
\label{fig:folding}
\end{center}
\end{figure}

In this work, the wave functions of the deuteron, $^3$He and $^3$H nuclei are calculated in the ab initio approach, using the A$v$18 potential \cite{av18} as realistic nuclear force, and the phenomenological one-pion exchange CP-odd nuclear force \cite{pvcpvhamiltonian3}.
The CP-odd potential is given by
\begin{eqnarray}
H_{P\hspace{-.35em}/\, T\hspace{-.5em}/\, }^\pi
& = &
\bigg\{ 
\bar{G}_{\pi}^{(0)}\,{\vc{\tau}}_{1}\cdot {\vc{\tau}}_{2}\, {\vc{\sigma}}_{-}
+\frac{1}{2} \bar{G}_{\pi}^{(1)}\,
( \tau_{+}^{z}\, {\vc{\sigma}}_{-} +\tau_{-}^{z}\,{\vc{\sigma}}_{+} )
+\bar{G}_{\pi}^{(2)}\, (3\tau_{1}^{z}\tau_{2}^{z}- {\vc{\tau}}_{1}\cdot {\vc{\tau}}_{2})\,{\vc{\sigma}}_{-} 
\bigg\}
\cdot
\hat{ \vc{r}} \,
V(r)
,
\label{eq:CPVhamiltonian}
\end{eqnarray}
where $\hat{\vc{r}} $ is the unit vector of the relative coordinate $\vc{r} \equiv \vc{r}_1 - \vc{r}_2$ with the subscripts denoting the interacting nucleons.
The spin and isospin matrices are given by ${\vc{\sigma}}_{-} \equiv {\vc{\sigma}}_1 -{\vc{\sigma}}_2$, ${\vc{\sigma}}_{+} \equiv {\vc{\sigma}}_1 + {\vc{\sigma}}_2$, ${\vc{\tau}}_{-} \equiv {\vc{\tau}}_1 -{\vc{\tau}}_2$, and ${\vc{\tau}}_{+} \equiv {\vc{\tau}}_1 + {\vc{\tau}}_2$.
The dimensionless CP-odd couplings $\bar G_\pi^{(i)}$ $(i=0,1,2)$ are small, and are considered as given.
The radial part of the CP-odd potential is given by
\begin{equation}
V(r)
= 
-\frac{m_\pi}{8\pi m_N} \frac{e^{-m_\pi r }}{r} \left( 1+ \frac{1}{m_\pi r} \right)
\ .
\end{equation}
Its radial dependence on $r$ is shown in Fig. \ref{fig:folding}.

\subsection{The $\alpha - N$ interaction}

The wave function of light nuclei is well described in the cluster model \cite{yamada,clusterreview3}.
In this study, the $\alpha$-cluster is considered as a degree of freedom for the $^6$Li, $^9$Be, and $^{13}$C nuclei.
For the CP-even $\alpha - N$ interaction, we adopt the phenomenological effective interactions which reproduce the phase shift of low energy scattering between light nuclei and nucleon \cite{kanada,hasegawa,schmid}.
The effect of antisymmetization is taken into account by using the Orgonality Condition Model \cite{saito} where the forbidden states are projected out.

The CP-odd $\alpha -N$ potential is given by folding the bare CP-odd $N-N$ interaction of Eq. (\ref{eq:CPVhamiltonian}).
After eliminating the center of mass motion \cite{horiuchi}, it is written as
\begin{equation}
V_{\alpha -N } 
( R ) 
\frac{\vc{R } }{|\vc{R }|}
=
\int d^3 \vc{R}' \,
V 
(|\vc{R } -\vc{R }' | ) \,
\rho_\alpha (R')
\frac{\vc{R } -\vc{R }'}{|\vc{R } -\vc{R }'|}
,
\end{equation}
where $\rho_\alpha ( R ) = \frac{32}{3\sqrt{3} \pi^{3/2} b^3} e^{-\frac{4}{3} R^2/b^2}$ with the oscillator constant $b=1.358$ fm, and $\vc{R}$ is the $\alpha -N$ relative coordinate.
The shape of the CP-odd $\alpha - N$ interaction is shown in Fig. \ref{fig:folding}.
The isoscalar and isotensor CP-odd nuclear forces cancel due to the closed spin and isospin shells of the $\alpha$-cluster.

\subsection{The Gaussian expansion method}

The Schr\"{o}dinger equation of the few-body systems can be accurately solved using the Gaussian Expansion Method \cite{hiyama}.
We have to solve the nonrelativistic Schr\"{o}dinger equation
\begin{eqnarray}
( H - E ) \, \Psi_{JM_z,TT_z}  = 0 
,
\label{eq:schr7}
\end{eqnarray}
by diagonalizing the hamiltonian $H$ in the Gaussian basis.
The Gaussian Expansion Method is a variational method, and a set of trial functions is required.
It is given as a superposition of Gaussian base functions
\begin{eqnarray}
\phi_{nlm}(\vc{r})
&=&
N_{nl } r^l \, e^{-(r/r_n)^2}
Y_{lm}({\widehat {\vc r}})  \;  ,
\end{eqnarray}
where $N_{nl} $ is the normalization constant.
The Gaussian range parameters follow the geometric progression
\begin{eqnarray}
r_n
&=&
r_1 a^{n-1} \qquad \enspace
(n=1 - n_{\rm max}) \; .
\end{eqnarray}
Using this trial function, it is possible to incorporate both the long distance physics as well as the short distance one.
To express the $A$-body wave functions, $A-1$ relative coordinates are required.
For convenience, we use the Jacobi coordinate to formulate them.

\section{The nuclear electric dipole moment}

The nuclear EDM is made of two leading contributions.
The first one is the intrinsic nucleon EDM contribution, given by
\begin{eqnarray}
d_A^{\rm (Nedm)} 
&=&
\sum_{i}^A
d_i \langle \, \Phi_J(A) \, |\, \sigma_{iz} \, |\, \Phi_J(A) \, \rangle
\equiv 
K_n^A d_n 
+K_p^A d_p 
,
\label{eq:nedmnuclearedm}
\end{eqnarray}
where $|\, \Phi_J(A) \, \rangle$ is the polarized nuclear wave function, $d_n$ and $d_p$ are the EDMs of the neutron and the proton, respectively.

The second contribution to the nuclear EDM is the polarization contribution, which is generated in the presence of the P, CP-odd nuclear force.
This effect is given by
\begin{eqnarray}
d_{A}^{\rm (pol)} 
&=&
\sum_{i=1}^{A} \frac{e}{2} 
\langle \, \tilde \Phi_J (A) \, |\, (1+\tau_i^3 ) \, r_{iz} \, | \, \tilde \Phi_J (A) \, \rangle
,
\label{eq:polarizationedm}
\end{eqnarray}
where $|\, \tilde \Phi_{J=1/2} \, \rangle$ is the polarized nuclear wave function with parity mxing.

\section{Results and analysis\label{sec:results}}

The results of the theoretical evaluations of the EDM of light nuclei are summarized in Table \ref{table:nuclearedm}.
We now analyze the data by sorting them according to the relevant physics.

\begin{table}[tbh]
\caption{
The EDM coefficients of the pion exchange CP-odd nuclear force.
The linear coefficients of the CP-odd nuclear couplings $a_\pi^{(i)}$ ($i=0,1,2 $) are expressed in unit of $10^{-2} e$ fm.
The sign $-$ means that the result vanishes in our setup.
The data of the neutron EDM is also shown for comparison \cite{crewther}.
}
\label{table:nuclearedm}
\begin{center}
\begin{tabular}{l|cc|ccc|}
  & $K_n$ & $K_p$ &$a_\pi^{(0)}$ & $a_\pi^{(1)}$ & $a_\pi^{(2)}$ \\ 
\hline
$n$ \cite{crewther} & 1 & 0 & $1$ & $- $ & $-1$  \\
$^{2}$H \cite{yamanakanuclearedm} & 0.914 & 0.914 & $-$ & $1.45 $ & $-$  \\
$^{3}$He \cite{yamanakanuclearedm} & 0.88 & $-0.04$ & $0.59$ & 1.08 & 1.68  \\
$^{3}$H \cite{yamanakanuclearedm} & $-0.05$ & 0.88 & $-0.59$ & 1.08 & $-1.70$  \\
$^{6}$Li \cite{yamanakanuclearedm} & 0.86 & 0.86 & $-$ & 2.2 & $-$  \\
$^{9}$Be \cite{yamanakanuclearedm} & 0.75 & $-$ & $-$ & $1.4$ & $-$  \\
$^{13}$C \cite{c13edm} & $-0.33$ & $-$ & $-$ & $-0.20 $ & $-$  \\
\hline
\end{tabular}
\end{center}
\end{table}

\subsection{Intrinsic nucleon EDM contribution}

As we can see in Table \ref{table:nuclearedm}, the linear coefficients  of the nucleon EDM contribution to the nuclear EDM ($K_n, K_p$) are less than one.
This result shows that this effect is not enhanced for light nuclei.
This is because the nucleons are not relativistic inside the nucleus, and we cannot expect relativistic enhancement as for the electron EDM effect inside atoms \cite{sandars}.
Rather, the nucleon EDM is suppressed due to the mixing of angular momentum configurations which acts destructively to the valence nucleon spin.
It was also recently pointed out that the polarization of the nuclear system by the EDM of nucleons may partially suppress the EDM of the nucleus, like the screening phenomenon of Schiff \cite{inoue}.

The nucleon EDM contribution is important for models of new physics which contribute through the quark EDM.
In that case, we have to note that the quark EDM contribution to the nucleon EDM is also suppressed by the nucleon tensor charge which collects the dynamical QCD effect to the quark EDM \cite{yamanakasde1,courtoy,kang}.
Recent lattice QCD data are giving $d_n \approx 0.8 d_d - 0.2 d_u$ \cite{etm3,bhattacharya2,jlqcd4}.
We must also note that the Wilson coefficient of the quark EDM operator is suppressed in the change of scale by the renormalization group evolution \cite{Dekens,degrassi} (typically, about 80\% when we run from $\mu =1$ TeV down to $\mu = 1 $ GeV).

\subsection{Polarization contribution to the EDM of the deuteron, $^3$He, and $^3$H}

The polarization contribution to the EDM of the deuteron, $^3$He, and $^3$H was evaluated ab initio using the Gaussian Expansion Method.
The result is consistent with the phenomenological and chiral effective field theory analyses \cite{liu,bsaisou}.
We must note that the deuteron has very weak sensitivity on the isoscalar and isotensor CP-odd nuclear forces, due to the isospin symmetry.
The EDMs of $^3$He and $^3$H are sensitive to all isospin structures.

The systematics due to the choice of the realistic 2-body nuclear force is almost irrelevant in the evaluation of the nuclear EDM \cite{afnan}. 
This shows that the EDM of light nuclei is well described by the long distance physics, which is accurately described by the pion exchange.
As potentially important sources of systematics, we have the effect of three-body force \cite{3bodyforce} and the exchange current contribution \cite{pastore}.
Those effects were found to be less than 10\% in previous works \cite{liu,stetcu}.
The effects of heavier meson exchange are also smaller by more than an order of magnitude \cite{yamanakanuclearedm}.

\subsection{Polarization contribution to the EDM of $^6$Li and $^9$Be}

The $^6$Li and $^9$Be nuclei were evaluated as three-body systems in the $\alpha$-cluster model.
As a result, it is found that the sensitivity of the EDM of $^6$Li on the isovector pion exchange CP-odd nuclear force is larger than that of other nuclei.
This enhancement is due to the constructive interference of the EDM of the deuteron subsystem inside $^6$Li and the $\alpha - N$ polarization.
We must note that $^6$Li has weak sensitivity on the isoscalar and isotensor CP-odd nuclear forces, since the deuteron and the $\alpha -N$ subsystems are protected by the isospin symmetry.
The EDM of $^9$Be has an $\alpha - \alpha$ subsystem which cannot be polarized, so we find a smaller EDM than for $^6$Li.
The $^9$Be nucleus is also only sensitive to the isovector CP-odd nuclear force.
Our result suggests that the cluster structure may enhance the EDM of nuclei.

\subsection{Polarization contribution to the EDM of $^{13}$C}

The polarization contribution to the EDM of $^{13}$C was evaluated in the four-body cluster model \cite{c13edm}.
From our calculation, it is found to be smaller than other nuclei studied.
The physical mechanism which suppresses the EDM of $^{13}$C is the small overlap between the $^{12}$C core of the $1/2_1^-$ and $1/2_1^+$ states.
It is known that the $1/2_1^-$ state of $^{13}$C (ground state in the absence of CP-odd nuclear force) has a $^{12}$C ($2^+$) core due to the strong spin-orbit force between the core and valence neutron, whereas the $1/2_1^+$ state has a neutron halo structure, with a $^{12}$C ($0^+$) core \cite{yamada}.
The parity violating transition between those states are suppressed so that the EDM of $^{13}$C is small.
This result shows that the nuclear EDM is very sensitive to the nuclear structure, and the na\"{i}ve expectation that the presence of opposite parity levels close to the ground state (3.1 MeV for $^{13}$C) does not always enhance it.

\section{Prospects of new physics discovery\label{sec:prospects}}

We now show the prospects for the observation of new physics beyond standard model through the nuclear EDM.
By modeling the new physics contribution by the tree level exchange of new particles with mass $M_{NP}$ with $O(1)$ CP phase,  the typical CP-odd nuclear coupling becomes $\bar G_\pi \sim g^2_{NP} \frac{\Lambda_{\rm QCD}^2}{M_{NP}^2} $, with $\Lambda_{\rm QCD} \sim 200$ MeV and the typical coupling constant $g_{NP}$.
If the EDM of nuclei studied in this work is measured at the level of $O(10^{-29})e$ cm, we can probe new physics with the energy scale $M_{NP} \sim$ PeV [with $g_{NP} = O(0.1)$].
This na\"{i}ve estimation is adequate for models which generate CP-odd 4-quark interactions, such as the Left-right Symmetric Model \cite{Dekens}.

For the supersymmetric model, the nuclear EDM can probe the CP phases of the supersymmetric couplings at the level of $O(10^{-2})$ for the typical supersymmetry breaking scale $M_{\rm SUSY} \sim$ TeV, if the sensitivity of $O(10^{-29})e$ cm is realized in its measurement \cite{pospelovreview}.
The experimental study of the nuclear EDM provides a complementary analysis to high energy accelerator based experiments, and may unveil the high scale supersymmetry breaking through CP violation.

We also give here the sensitivity of the nuclear EDM to the class of models which generate Barr-Zee type diagrams.
This is the case for the Higgs doublet models \cite{2higgs}, supersymmetric models without R-parity \cite{yamanakabook,rpvlinearprogramming}, or the minimal CP violating model with the new 750 GeV boson suggested with high statistics by the LHC experimental data \cite{750,goertz,choi}.
With a simple dimensional analysis, the sensitivity on the energy scale of new physics is $M_{NP} \sim \sqrt{Y_q Y_Q} $ PeV, with the coupling constants $Y_q$ and $Y_Q$ which attenuate it, if they are small.

\section{Summary\label{sec:summary}}

In this work, we have calculated the EDM of the deuteron, $^3$He, $^3$H, $^6$Li, $^9$Be, and $^{13}$C using the Gaussian expansion method.
The deuteron, $^3$He, and $^3$H were calculated ab initio, and $^6$Li, $^9$Be, and $^{13}$C was evaluated in the $\alpha$-cluster model.

In the ab initio approach, the results found are consistent with other works.
For the $^6$Li, we have found an enhancement of the sensitivity on the isovector pion exchange CP-odd nuclear force.
For the EDM of $^{13}$C, we have obtained a smaller enhancement factor than those of other nuclei evaluated in this work.
The result of our work also suggests that several nuclei such as $^7$Li or $^{19}$F are sensitive to the nucleon level CP violation.
The theoretical evaluation of them is our next target.

The sensitivity of the nuclear EDM on the new physics beyond standard model can reach the level of PeV, which is well beyond the experimental sensitivity of LHC, and we recommend the realization of the nuclear EDM experiments.

\end{document}